\documentclass[twocolumn,prd,nofootinbib]{revtex4}

\newcommand\ForInternalReference[1]{}
\newcommand\SkipForEarlyCirculation[1]{}
\newcommand\AddedResponse[1]{ {#1}}
\newcommand\SkipPP[1]{}
\usepackage{verbatim}
\usepackage{graphicx}
\usepackage{dcolumn}
\usepackage{bm}
\usepackage{color}
\usepackage{xspace}
\usepackage{url}
\usepackage{amsmath}
\usepackage{float}
\usepackage{multirow}
\usepackage{times}
\newcommand\optional[1]{}

\newcommand\qmstateproduct[2]{\left\langle#1|#2\right\rangle}

\newcommand\Y[1]{{{}_{#1}Y}}

\newcommand\lnLmarg{ \ln{\cal L}_{\rm marg}{}}
\newcommand\unit[1]{{\rm #1}}

\newcommand\ILE{ILE}
\newcommand\editremark[1]{{\color{red} #1}}
\usepackage{color}
\definecolor{amber}{rgb}{1.0, 0.75, 0.0}
\definecolor{orange}{rgb}{1.0, 0.5, 0.0}
\definecolor{amaranth}{rgb}{0.9, 0.17, 0.31}

\graphicspath{{./figures/}}
\newcommand{\mc}{{\cal M}}

\def\ltsima{$\; \buildrel < \over \sim \;$}
\def\simlt{\lower.5ex\hbox{\ltsima}}
\def\gtsima{$\; \buildrel > \over \sim \;$}
\def\simgt{\lower.5ex\hbox{\gtsima}}

\def\RIT{Center for Computational Relativity and Gravitation, Rochester Institute of Technology, Rochester, New York 14623, USA}

\begin{document}

\title{Accelerating parameter inference with graphics processing units}
\author{D. Wysocki}
\affiliation{\RIT}
\author{R. O'Shaughnessy}
\affiliation{\RIT}
\author{Yao-Lung L. Fang}
\affiliation{Computational Science Initiative, Brookhaven National Laboratory, Upton, NY 11973, USA}
\affiliation{National Synchrotron Light Source II, Brookhaven National Laboratory, Upton, NY 11973, USA}
\author{Jacob Lange}  %
\affiliation{\RIT}
\begin{abstract}
Gravitational wave Bayesian parameter inference involves repeated comparisons of GW data to generic candidate predictions.
Even with algorithmically efficient methods like RIFT or reduced-order quadrature, the time needed to perform these
calculations and overall computational cost can be significant compared to the minutes to hours needed to achieve the
goals of low-latency multimessenger astronomy.  By translating some elements of the RIFT algorithm to operate on
graphics processing units (GPU), we demonstrate substantial performance improvements, enabling dramatically reduced
overall cost and latency. 
\end{abstract}
\maketitle

\section{Introduction}
The Advanced LIGO  \cite{2015CQGra..32g4001T}  and Virgo \cite{gw-detectors-Virgo-original-preferred}  ground-based gravitational wave (GW) detectors have
identified several coalescing compact binaries
\cite{DiscoveryPaper,LIGO-O1-BBH,2017PhRvL.118v1101A,LIGO-GW170814,LIGO-GW170608,LIGO-GW170817-bns}.  
The properties of the sources responsible have been inferred via Bayesian comparison of  each source with candidate
gravitational wave signals
\cite{DiscoveryPaper,LIGO-O1-BBH,2017PhRvL.118v1101A,LIGO-GW170814,LIGO-GW170608,LIGO-GW170817-bns,LIGO-O2-Catalog,gwastro-PE-AlternativeArchitectures,gwastro-PENR-RIFT,gw-astro-PE-lalinference-v1}.
Many more are expected as observatories reach design sensitivity \cite{2016LRR....19....1A}.  
Both to handle the rapid pace of discovery and particularly to enable coordinated multimessenger followup, GW
observatories should be able to reconstruct the evidence for a signal being present in the data along with the full source parameters of coalescing binaries as fast as possible
\cite{LIGO-2016-WhitePaper,LIGO-Program-2018}.
However, particularly when using the best-available waveform models, these calculations can be very costly.  For binary
neutron stars, detailed inferences even using simplified waveforms take weeks.  Even for massive binary  black holes,
these calculations can take  months for costly time-domain effective-one-body models which incorporate the effects of precession.

Several strategies have been
developed to reduce the 
computational cost of parameter estimation \cite{gwastro-pe-Brandon-STF2,gwastro-PE-AlternativeArchitectures,gw-astro-ReducedOrderQuadraturePE-TiglioEtAl2014,2016PhRvD..94d4031S,2017CQGra..34k5006V,2018arXiv180608792Z}.  
Approaches that have appeared in the literature include
generating the approximate solutions more quickly \cite{gwastro-mergers-PE-ReducedOrder-2013,2014PhRvX...4c1006F,2013PhRvD..87l2002S,2013PhRvD..87d4008C,gwastro-mergers-IMRPhenomP,gwastro-SpinTaylorF2-2013}; 
interpolating some combination
of the waveform or likelihood \cite{gwastro-approx-ROMNR-Blackman2015,2013PhRvD..87d4008C,2013PhRvD..87l2002S,2014CQGra..31s5010P,2014PhRvD..90d4074S,gwastro-PE-AlternativeArchitectures,cole2014likelihood,2012MNRAS.421..169G};  or 
adopting a sparse representation to reduce the computational cost of data
handling \cite{antil2013two,gwastro-mergers-PE-ReducedOrder-2013,2016PhRvD..94d4031S,gw-astro-ReducedOrderQuadraturePE-TiglioEtAl2014,gwastro-PE-AlternativeArchitectures,2010arXiv1007.4820C,2018arXiv180608792Z}.   %
\AddedResponse{Not published but also important are code optimization projects that improve    infrastructure,
  such as better parallelization for \texttt{lalinference} \cite{gw-astro-PE-lalinference-v1,lalsuite}.}
 Some methods, however, achieve rapid turnaround through simplifying approximations.  
The RIFT/rapidPE strategy described in \cite{gwastro-PE-AlternativeArchitectures,2017CQGra..34n4002O,gwastro-PENR-RIFT,NRPaper} eschews these simplifications, performing
embarrassingly-parallel inferences even for costly models.   However, the method as previously implemented still had a
significant net  computational cost and noticeable startup time, limiting its diverse potential applications.
Previously, B. Miller  developed a custom-coded  implementation of the
relevant likelihood on graphics processing units (GPUs), suggesting substantial performance improvements were possible
\cite{Thesis-Brandon-2016,GPU-NU-work}.  
\AddedResponse{The pycbc search code\cite{2016CQGra..33u5004U} and its pycbc-inference extension
  \cite{2019PASP..131b4503B} can also make use of several GPU
  hardware-accelerated operations.
}
In this paper,  we describe a variant of the RIFT approach which flexibly translates one of its
algorithms to operate on GPUs, and adjusts its workflow to exploit the speed improvements
this re-implementation affords.

This paper is organized as follows.
In Section \ref{sec:methods}, we review the underlying marginalized likelihood calculations used by RIFT and  their updated GPU
implementation.  
In Section \ref{sec:demo}, we quantify the improved performance of our  GPU-accelerated code,   while assessing
operating settings which facilitate increased performance.
In Section \ref{sec:end-to-end}, we describe the performance of the end-to-end pipeline on the synthetic nonspinning signals used in
Section \ref{sec:demo}.  
In Section \ref{sec:conclude}, we summarize our results and discuss their potential applications to future GW source and
population inference.

\section{Methods}
\label{sec:methods}

\subsection{Parameter inference with the RIFT likelihood}
\ILE{}  -- a specific algorithm to ``Integrate (the Likeilhood) over Extrinsic parameters'' -- provides a straightforward and efficient mechanism to compare any specific candidate gravitational wave source with
real or synthetic data   \cite{gwastro-PE-AlternativeArchitectures,NRPaper,2017PhRvD..96j4041L,2017CQGra..34n4002O},
by marginalizing the likelihood of the data over the seven coordinates characterizing the spacetime coordinates and
orientation of the binary relative to the earth.  
Specifically the likelihood of the data given gaussian noise, relative to gaussian noise, has the form  (up to normalization)
\begin{equation}
\label{eq:lnL}
\ln {\cal L}(\bm{\lambda} ,\theta )=-\frac{1}{2}\sum\limits_{k}\langle h_{k}(\bm{\lambda} ,\theta )-d_{k} |h_{k}(\bm{\lambda} ,\theta )-d_{k}\rangle _{k}-\langle d_{k}|d_{k}\rangle _{k},
\end{equation}
where $h_{k}$ are the predicted response of the k$^{th}$ detector due to a source with parameters ($\bm{\lambda}$, $\theta$) and
$d_{k}$ are the detector data in each instrument k; $\bm{\lambda}$ denotes the combination of redshifted mass $M_{z}$ and the
remaining parameters needed to uniquely specify the binary's dynamics; $\theta$ represents the
seven extrinsic parameters (4 spacetime coordinates for the coalescence event and 3 Euler angles for the binary's
orientation relative to the Earth); and $\langle a|b\rangle_{k}\equiv
\int_{-\infty}^{\infty}2df\tilde{a}(f)^{*}\tilde{b}(f)/S_{h,k}(|f|)$ is an inner product implied by the k$^{th}$ detector's
noise power spectrum $S_{h,k}(f)$. 
In practice we adopt both  low- and high- frequency cutoffs $f_{\rm max},f_{\rm min}$ so all inner products are modified to
\begin{equation}
\label{eq:overlap}
\langle a|b\rangle_{k}\equiv 2 \int_{|f|>f_{\rm min},|f|<f_{\rm max}}df\frac{[\tilde{a}(f)]^{*}\tilde{b}(f)}{S_{h,k}(|f|)}.
\end{equation}
The joint posterior probability of $\bm{\lambda} ,\theta$ follows from Bayes' theorem:
\begin{equation}
p_{\rm post}(\bm{\lambda} ,\theta)=\frac{ {\cal L}(\bm{\lambda} ,\theta)p(\theta)p(\bm{\lambda})}{\int d\bm{\lambda} d\theta {\cal L}(\bm{\lambda} ,\theta)p(\bm{\lambda})p(\theta)},
\end{equation}
where $p(\theta)$ and $p(\bm{\lambda})$ are priors on the (independent) variables $\theta ,\bm{\lambda}$. For each $\bm{\lambda}$, we evaluate the marginalized likelihood
\begin{equation}
\label{eq:lnLmarg}
 {\cal L}_{\rm marg}\equiv\int  {\cal L}(\bm{\lambda} ,\theta )p(\theta )d\theta
\end{equation}
via direct Monte Carlo integration over almost all parameters $\theta$, where $p(\theta)$ is uniform in 4-volume and source orientation.  
For the event time parameter, we marginalize by direct quadrature, for each choice of the remaining Monte Carlo parameters.
For the remaining dimensions, to evaluate the likelihood in regions of high importance, we use an adaptive Monte Carlo as described in
\cite{gwastro-PE-AlternativeArchitectures}.

This marginalized likelihood can be evaluated efficiently
by  generating the dynamics and outgoing radiation in all possible directions once and for all for fixed
$\bm{\lambda}$, using a spherical harmonic decomposition.  Using this cached information effectively,  the likelihood can be evaluated as a function of $\theta$ at very low computational cost.  
A dimensionless, complex gravitational-wave
strain
\begin{align} \label{eq:strain}
h(t,\vartheta,\phi;\bm{\lambda}) =  h_+(t,\vartheta,\phi;\bm{\lambda}) - 
                                i h_\times (t,\vartheta,\phi;\bm{\lambda}) \, ,
\end{align}
can be expressed in terms of its two fundamental polarizations $h_+$ and $h_\times$.
Here, $t$ denotes time, $\vartheta$ and $\phi$ are the polar and azimuthal angles
for the direction of gravitational wave propagation away from the source. 
The complex gravitational-wave strain can be written in terms of
spin-weighted spherical harmonics $\Y{-2}_{\ell m} \left(\vartheta, \phi \right)$ as 
\begin{align} \label{eq:strain_mode}
h(t,\vartheta,\phi;\bm{\lambda}) = 
\sum_{\ell=2}^{\infty} \sum_{m=-\ell}^{\ell} \frac{D_{\rm ref}}{D} h^{\ell m}(t;\bm{\lambda}) \Y{-2}_{\ell m} \left(\vartheta, \phi \right) \, ,
\end{align}
where the sum includes all available harmonic modes $h^{\ell m}(t;\pmb{\bm{\lambda}})$ made available by the model;  where
$D_{\rm ref}$ is a fiducial reference distance; and where $D$, the luminosity distance to the  source, is one of the
extrinsic parameters.  

Following \citet{gwastro-PE-AlternativeArchitectures}, we substitute expression~\eqref{eq:strain_mode} 
for $h_{\ell m}$ into the expression $h_k(t) =F_{+,k} h_+(t_k) +
  F_{\times,k}h_\times(t_k)$ for the detector response $h_k$, 
where $t_k=t_c - \vec{x}_k \cdot \hat{n}$ is the arrival time at the $k$th detector (at position $\vec{x}_k$)
for a plane wave propagating along $\hat{n}$ \cite{gwastro-PE-AlternativeArchitectures}.
We then substitute these expressions for $h_k$ into the likelihood function~\eqref{eq:lnL}
thereby generating~\cite{gwastro-PE-AlternativeArchitectures}
\begin{widetext}
\begin{align}
\ln {\cal L}(\bm{\lambda}, \theta) 
&= (D_{\rm ref}/D) \text{Re} \sum_k \sum_{\ell m}(F_k \Y{-2}_{\ell m})^* Q_{k,lm}(\bm{\lambda},t_k)\nonumber \\
&   -\frac{(D_{\rm ref}/D)^2}{4}\sum_k \sum_{\ell m \ell' m'}
\left[
{
|F_k|^2 [\Y{-2}_{\ell m}]^*\Y{-2}_{\ell'm'} U_{k,\ell m,\ell' m'}(\bm{\lambda})
}
 {
+  \text{Re} \left( F_k^2 \Y{-2}_{\ell m} \Y{-2}_{\ell'm'} V_{k,\ell m,\ell'm'} \right)
}
\right]
\label{eq:def:lnL:Decomposed}
\end{align}
\end{widetext}
where 
 where $F_k = F_{+,k} - i F_{\times,k}$ are the
complex-valued detector
response functions of the $k$th detector \cite{gwastro-PE-AlternativeArchitectures} and 
the quantities $Q,U,V$ depend on $h$ and the data as
\begin{subequations}
\label{eq:QUV}
\begin{align}
Q_{k,\ell m}(\bm{\lambda},t_k) &\equiv \qmstateproduct{h_{\ell m}(\bm{\lambda},t_k)}{d}_k \nonumber\\
&= 2 \int_{|f|>f_{\rm low}} \frac{df}{S_{n,k}(|f|)} e^{2\pi i f t_k} \tilde{h}_{\ell m}^*(\bm{\lambda};f) \tilde{d}(f)\ , \\
{ U_{k,\ell m,\ell' m'}(\bm{\lambda})}& = \qmstateproduct{h_{\ell m}}{h_{\ell'm'}}_k\ , \\
V_{k,\ell m,\ell' m'}(\bm{\lambda})& = \qmstateproduct{h_{\ell m}^*}{h_{\ell'm'}}_k  \ .
\end{align}
\end{subequations}

\AddedResponse{In RIFT, the marginalization in Eq. (\ref{eq:lnLmarg}) over extrinsic parameters is performed by evaluating this likelihood $\ln {\cal
    L}(\bm{\lambda},\theta)$ on long
  arrays $\theta_\alpha$ of extrinsic parameters, including $D_\alpha$ but excluding time.    Treating the block of quantities that arise
  in one sequence of evaluations together,  the expressions $Q,F,U,V$ and $D$ can be considered as multi-dimensional
  arrays.  For example, $F_k(\theta_\alpha)$ is a matrix indexed over $k$ (detectors) and $\alpha$
  (extrinsic parameters); $\Y{-2}_{lm}(\theta_k)$ is a matrix of shape (modes)$\times$(extrinsic); and $D_\alpha$ is a one-dimensional array over $\alpha$.  By contrast, the  $U,V$ are arrays of shape
  (detectors)$\times$ (modes)$^2$ are independent of extrinsic parameters.  While $Q$ depends explicitly on time, and
  varies rapidly on short timescales, $F$ varies on the earth's rotation timescale, so we  approximate $F$ as  time-independent. }

At a fixed set of intrinsic parameters $\bm{\lambda}$, \ILE{} will repeatedly evaluate the time-marginalized likelihood 
for each candidate Monte Carlo set of extrinsic parameters $\theta$:
\begin{eqnarray}
{\cal L}_{\rm margT} \equiv  \int {\cal L} \frac{dt}{T}
\label{eq:lnL:tmarg}
\end{eqnarray}
where $T$ is a small coincidence window consistent with the candidate event time; here and previously,
$T=0.3\unit{sec}$.  
\AddedResponse{
Because we treat the  parameter  asymmetrically when marginalizing over all extrinsic parameters, we likewise organize
a multidimensional array representation of $Q$ to emphasize the special role of time: for each $k,l,m$ we construct a uniformly-sampled timeseries
$Q_{k,lm}(\bm{\lambda},\tau)$ versus $\tau$, truncating it to a narrow coincidence window.   In effect, we represent $Q$
by a matrix 
  $Q_{k,lm}(\bm{\lambda},t+\hat{n}\cdot x_k)$ with  shape (detectors)$\times$(modes)$\times$(extrinsic)$\times$(time).
}
\AddedResponse{In terms of these multidimensional arrays, r}ewriting sums as matrix operations, the likelihood can be equivalently expressed as
\begin{align}
\ln {\cal L} &= \frac{D_{\rm ref}}{D} \text{Re}[ (F Y)^\dag Q]  \nonumber \\
 & - \frac{D_{\rm ref}^2}{4 D^2} [ (FY)^\dag U FY + (FY)^TV FY]
\label{eq:lnL:MatrixForm}
\end{align}
\AddedResponse{where this symbolic expression employs an implicit index-summation convention such that all naturally paired
  indices are contracted.  The result is an array of shape (time)$\times$(extrinsic).}

\subsection{Accelerated evaluation via efficient multiplication}
  To accelerate the code, after precomputing the inner products $U,V,Q$, we simply shift them to the
graphics card, then carry out all calculations necessary to implement Eqs. (\ref{eq:lnL:MatrixForm}, \ref{eq:lnL:tmarg})
on the GPU.  
\AddedResponse{These arrays are only a few kilobytes.  We then construct blocks of $10^4$ random extrinsic parameters $\theta_\alpha$
  with the CPU; transfer them to the board; and use on-board code to construct $10^4$ values for $\lnLmarg$. }
To enable this implementation with portable code, we use \texttt{cupy} \cite{cupy_learningsys2017}, a drop-in-replacement for equivalent
\texttt{numpy} code used for the CPU-based version of ILE.    For the most costly part of the calculation -- the inner
products necessary to evaluate $Q_{lm}(t)$, accounting for distinct time-of-flight-dependent time windows for each
interferometer's data -- we use a custom CUDA kernel to perform the necessary matrix multiplication.  
With these changes alone, the likelihood evaluation is roughly $60\times$ faster on equivalent hardware.  
After this update, individual likelihood evaluation costs are not the performance- or cost-limiting feature of RIFT-based source
parameter inference.  Instead, the overhead associated with the adaptive Monte Carlo and with the (CPU-based) inner
product evaluations for $Q,U,V$ dominate our computational cost.

\subsection{Tradeoff between Monte Carlo integration and accuracy}

Each marginalized likelihood evaluation has a relative uncertainty of order  $1/\sqrt{n_{\rm eff}}$, where the number of
effective samples $n_{\rm eff}=\epsilon N_{\rm it}$ increases linearly with the total number $N_{\rm it}$ of Monte Carlo
evaluations performed by \ILE{}.  Therefore, to increase (decrease) our accuracy for each likelihood evaluation by a factor of $A$
will require a factor $A^2$ more (fewer) iterations.   
We adopt a fixed threshold on $n_{\rm eff}$.

We do benefit slightly by re-using the adaptive Monte Carlo integrator for each extrinsic point $\bm{\lambda}$.  Since the
integrator has already identified the likely range of sky locations and distances on the first iteration, each
subsequent evaluation of the marginalized likelihood can converge marginally more quickly.

\section{Analysis of marginalized likelihood evaluation cost}
\label{sec:demo}

We illustrate the code using two synthetic signals: a binary black hole with masses $m_{1,2}=35,30 M_\odot$ and a binary
neutron star with masses $1.4 M_\odot$ and $1.35 M_\odot$.
 We perform this analysis on heterogeneous LVC collaboration computing resources \AddedResponse{described in more detail
   in the Appendix}, to assess our variable
performance across architectures.  Notably, we investigate the following GPU options: (a) GTX 1050 Ti at LIGO-CIT,
available in large numbers; (b) V100, available on selected high-performance machines; (c) and
GTX 1050 at LIGO-LHO.    Unless otherwise noted in the text, we will discuss code configurations using CPU-only and GPU (b) using
$4096\unit{Hz}$ sampling and only the  $\ell=2, m=\pm 2$ modes.  In this section, we  conservatively report computational cost for
parameters $\bm{\lambda}$ which are close to or within the posterior.  Because of their consistency with the data, 
the marginalized likelihood calculations converge the most slowly, as they have the best-determined extrinsic
parameters.

\subsection{Synthetic source generation and analysis settings}

We generate  synthetic data for a two-detector LIGO configuration, assuming operation at  initial (BBH) or advanced (BNS) design  sensitivity.
Both signals are generated with SEOBNRv4 effective-one-body waveform approximation \cite{2017PhRvD..95d4028B}, with a zero-noise data realization.   
To qualitatively reproduce the noise power spectrum and amplitude of typical binary black hole observations in O1 and
O2, the binary black hole signal has source distance $200 \unit{Mpc}$, chosen so SNR $\simeq 14$.  
Similarly, to qualitatively reproduce the analysis of GW170817, our synthetic binary neutron star is placed $100
\unit{Mpc}$ away, so the signal amplitude is SNR $\simeq 31$.

 We perform a multi-stage iterative RIFT analysis of these signal with the
time-domain SEOBNRv4 (BBH) or TaylorT4 \cite{gw-astro-PN-Comparison-AlessandraSathya2009}( BNS) approximation, under the simplifying assumption that both objects are point particles with zero
spin.   For both sources, we adopt the fiducial distance prior $p(D_L) \propto D_L^2$,   selecting  a maximum distance
roughly five times larger than the known source distance: 1Gpc for the binary black hole and 500 Mpc for the neutron
star.  We adopt a uniform prior on the component (redshifted) masses $m_{i,z}$ (and, when appropriate, spins
$\chi_{i,z}$).
Unless otherwise noted, all mass quantities described in this work are redshifted masses $m_{k,z}=(1+z)m_k$.
For the binary black hole we generate the signal with $f_{\rm min}=8 \unit{Hz}$ and analyze the signal with  $f_{\rm
  min}=10 \unit{Hz}$; for the binary neutron star, we generate the signal with $f_{\rm min}=20\unit{Hz}$ and anayze it
with $f_{\rm min}=23\unit{Hz}$.

\subsection{Massive compact binary black holes}
Previously, each instance of ILE examined one intrinsic point $\bm{\lambda}$.  The overall cost of this ILE evaluation
involved two parts: a startup cost, a setup cost, and a Monte Carlo integration cost.  The  startup cost $\tau_{start}$
is associated with code setup followed by reading, data conditioning, and Fourier-transforming the
data.  The setup cost $\tau_{setup}$ arises from waveform generation and  the inner products $U,V,Q$.  Finally, the Monte Carlo
cost $\tau_{mc}=N_{ad} \tau_{ad}+N_{it}\tau_{eval}$ increases with the number of Monte Carlo iterations $N_{it}$ with
and the number of times $N_{ad}$ the sampling prior used in adaptive Monte Carlo is regenerated from the most recent 
$n_{\rm chunk}$ data points, in proportion to their
cost.   In the standard configuration, the adaptive sampling prior is regenerated every $n_{\rm chunk}$ iterations (i.e,
 $N_{ad} \simeq N_{it}/n_{chunk}$), regenerating a sampling prior in two sky location parameters and distance.  
The choice of one $\bm{\lambda}$ for each instance of ILE was due to the substantial time  $\tau_{MC}$, which vastly
dominated the overall computational cost.   For example, for a typical analysis of a short signal --  a typical binary
black hole signal with $f_{\rm min}=20 \unit{Hz}$ and $m_1\simeq m_2\simeq 30 M_\odot$ --  this version has
cost elements as shown in Table \ref{tab:CostBreakdown}, based on the assumption that the MC terminates after $N_{\rm it}\simeq 2\times 10^6$ iterations using
$n_{\rm chunk}\simeq 10^4$.  
The marginalized likelihood described above  converges incredibly rapidly to a small value if the candidate model is
inconsistent with a signal (or the absence thereof) in the data, with only $N_{it} \simeq O(10^4)$  evaluations needed.

\begin{table*}
\begin{tabular}{lrr|ccccc|rr}
Version & srate & modes & $\tau_{start}$ & $\tau_{setup}$ & $\tau_{ad}$ & $\tau_{it,like}$ &$\tau_{it,rest}$ &
$\frac{T_{ILE}}{N_{eval}}$ & GPU \\  %
  &   Hz & m & sec & sec & & $\mu$sec & $\mu$sec  &sec  & use  \%\\ \hline 
CPU & 16384 & $\pm 2 $ & 20 & 2.4 &&540 & 20 &  690  \\ 
       & 4096 & $\pm 2 $ &   20  &&&& 20 \\ \hline
CPU & 16384 & $\pm 2,\pm 1 $ & 20 & 1.5 && 680 & 20 &  1060  \\ 
       & 4096 & $ \pm 2, \pm 1 $ &   20 &&&& 20  \\ \hline
GPU (a) & 16384 & $\pm 2 $  & 20  & & && & 270 \\
            & 4096 &$\pm 2 $  &  20 &  & & & & 45 \\ \hline
GPU (b) & 16384 & $\pm 2$ & 20  & 1.8 & 1 & 0.85& 20 &28 & 15\\
       & 4096 & $\pm 2$  & 20 & $1.2 $ &  1  & 0.75 & 20  & 25\\ \hline
GPU (b) & 16384 & $\pm 2, \pm 1$ & 20 & 1 && 4.2 & 20  & 38  \\
       & 4096 & $\pm 2, \pm 1$ & 20 & 1&& 2.5  & 20 & 35 & \\ \hline
GPU (c) & 16384 & $\pm 2 $  & 20  &6  & & 18&  58&160 &  \\
            & 4096 &$\pm 2 $  &  20 & 3.7 & & 11  & 58  & 140 & $\simeq 50$ \\
\end{tabular}
\caption{\label{tab:CostBreakdown}\textbf{Profiling performance: Binary black holes}: Evaluation costs for the
  marginalized likelihood on default
  hardware, for a two-mode system $(l,m)=\pm 2$ analyzing $T=8\unit{s}$ of data with a massive binary black hole
  $m_1=35 M_\odot,M_2=30 M_\odot$.  The last column indicates peak GPU utilization.
}
\end{table*}

With the new low-cost ${\cal L}_{\rm margT}$ evaluations, however, the startup and setup costs  $\tau_{start},\tau_{setup}$ now can be
comparable or in excess of the total time used to evaluate the marginalized likelihood ${\cal L}_{marg}$.   In fact, the total time per Monte Carlo
evaluation $\tau_{it}=\tau_{manage}+\tau_{eval}$  spent in general-purpose overhead $\tau_{manage}$ will be much larger than the time $\tau_{eval}$ spent
carrying out scientific calculations by evaluating the likelihood.  For these reasons, we reorganize the workflow, so each
instance of ILE loops over
$N_{eval}$ different choices for $\bm{\lambda}$. 
Additionally, particularly for massive binary black holes, we investigate two additional performance improvements.  First and
foremost, we lower the sampling rate, thus lowering the startup cost $\tau_{start}$ and particularly setup cost
$\tau_{setup}$.  Previously and out of an abundance of caution, \ILE{} employed a sampling rate $1/\Delta t = 16384
\unit{Hz}$ for all calculations, but terminated all inner products in Eq. (\ref{eq:overlap}) at roughly $f_{\rm
  max}=2048\unit{Hz}$ or less.   Reducing the sampling rate by a factor $s$ will \AddedResponse{reduce} the  cost of all operations with
timeseries -- they are shorter.  Depending on the relative cost of overhead versus array operations,  $\tau_{setup}$
\emph{and} the cost per evaluation $\tau_{eval}$ decrease modestly because of this effect.  
Also following Pankow et al, to insure safe inner products over short data sets in the presence of very narrow spectral lines, we operated on a
significantly-padded data buffer and modified the noise power spectrum by truncating the inverse power spectrum to a
finite response time.  For binary black holes, the degree of padding was significantly
in excess of the signal duration.  
Instead and following lalinference \cite{gw-astro-PE-lalinference-v1}, we adopt a much shorter inverse spectrum truncation length, a much shorter padding buffer, and
a Tukey window applied to the data to remove discontinuities at the start and end of each segment.   
Reducing the duration of data analyzed by a factor $s'$ will reduce the cost $\tau_{setup}$ by a factor $s'$.  
Combining these factors together, the overall computational cost  $T_{ILE}/N_{eval}$  of each marginalized likelihood
evaluation is
\begin{align}
T_{ILE}/N_{\rm eval} &= \frac{\tau_{start}}{N_{\rm eval}} 
 + 
 [ \tau_{setup} + N_{ad} \tau_{ad} + N_{it}\tau_{it}
 ] 
\end{align}
The first rows in table Table  \ref{tab:CostBreakdown} shows our estimated breakdown of these elements of the computational cost, for a typical binary black hole signal with $f_{\rm min}=20 \unit{Hz}$ and $m_1\simeq m_2\simeq 30 M_\odot$.
Because of the extremely high cost of adaptive overhead, we only adapt in two parameters, corresponding to sky location; 
and only adapt until a marginalied likelihood evaluation  returns significant $\ln {\cal L}_{\rm marg}$.  As a result, $N_{ad}$
(the number of adaptive stages \emph{per likelihood evaluation}) scales as $N_{it}/n_{chunk}/N_{\rm eval}$ and does not
increase with the overall number of samples, so the overall evaluation time scales as
\begin{align}
T_{ILE,mod}/N_{\rm eval} &= \frac{\tau_{start} + \tau_{ad}N_{it}/n_{\rm chunk}}{N_{\rm eval}} 
 \nonumber \\ &
 + 
 [ \tau_{setup}  + N_{it}\tau_{it}
 ]  \\
& \simeq (\tau_{\rm start}/20\unit{s}) +  10 (\tau_{ad}/\mu s) 
 \nonumber \\ &
+ 2  [(\tau_{it,liike}/\mu s) + \tau_{it,rest}/\mu s]
\end{align}
where in the latter expression we substitute the usual values of $N_{it} \simeq 2\times 10^6$,  $n_{\rm chunk}\simeq
10^4$, and $N_{\rm eval}\simeq 20$ conservtively adopted
for problems involving many degrees of freedom $d$ (between 6 to 8).  We target $N_{\rm eval} \simeq 20$ so that startup and other one-time costs  are not a significant contributor to the overall
runtime.   In this configuration, the run time per marginalized likelihood is around $T_{ILE}/N_{\rm eval} \simeq 25
\unit{s}$ for a binary black hole.    Of this $25 \unit{s}$, only roughly
$1.5\unit{s}$ corresponds to evaluating the likelihood, implying the cost $\tau_{it} \simeq 1.5/N_{\rm it} \simeq
7.5\times 10^{-7}$.
Likewise, because overhead dominates over array manipulations, an analysis with  $16\unit{kHz}$ timeseries requires only
marginally more time than the corresponding $4\unit{kHz}$ analysis.
Conversely, the overhead of  performing the Monte Carlo (including both $\tau_{ad}$ and $\tau_{it,rest}$) is a quite substantial contribution to the overall runtime for the best-available hardware.

For low-latency analyses, we need to assess the overall wallclock time needed to perform an analysis, often with a
restricted number $N_{GPU}$ of available GPU-enabled machines.   To  complete $N_{\rm net}$ likelihood evaluations
with these resources requires
\begin{align}
T_{net,mod} &= \frac{N_{\rm net}}{N_{\rm GPU} N_{\rm eval}} T_{ILE,mod} \nonumber \\
& =
  \frac{N_{net}}{N_{GPU}}\left[
\frac{\tau_{start} + \tau_{ad}N_{it}/n_{\rm chunk}}{N_{\rm eval}} 
 \right. \nonumber \\
& \left. + 
 [ \tau_{setup}  + N_{it}\tau_{it}
 ] 
 \right]
\end{align}
Typically we target $N_{\rm eval}  \lesssim 3\times 10^4, N_{\rm GPU} \simeq 100$, or roughly $300\times$ the cost per
individual marginalized likelihood evaluation.  For the binary black hole configuration described above, marginalized
likelihood evaluations will complete in about $100\; \unit{minutes}$ on a cluster of (b), or more realistically 200
minutes on a cluster of (a).  This number can be reduced to
tens of minutes with a modestly larger GPU pool, or by  marginally more conservative convergence thresholds
 $n_{\rm   eff}$ or $N_{it}$.  
This discussion ignores the latency introduced by the  gaussian process interpolation stage at the end.  We will revisit
accelerated GP interpolation in subsequent work.

The \ILE{} likelihood  generates waveform dynamics and $h_{lm}(t)$ once per evaluation $\bm{\lambda}$.   Waveform generation can contribute significantly to the
time needed to evaluate  $\ln {\cal L}_{\rm marg}$ for extremely costly
waveform models which require $\tau_{wf} $ a significant fraction of $  \tau_{setup}$.   That said, the tests described
above used relatively costly waveform generation with $h_{lm}(t)$ allocation requiring $\tau_{wf} \simeq 1.2 \unit{s}$
at 4kHz and $1.9\unit{s}$ at 16 kHz.

The \ILE{} likelihood $\ln {\cal L}_{\rm margT}$ involves sums over modes, with the most expensive likelihood and setup operations (involving $Q$) growing linearly with
the number of modes used.  This increased cost is most apparent in Table \ref{tab:CostBreakdown} when comparing the
columns corresponding  $\tau_{it,like}$ for GPU (b) between the $\pm 2$ and $(\pm 2,\pm 1)$ rows.  The cost of
evaluating $\ln {\cal L}_{\rm margT}$   with 4 modes is roughly twice that of runs with two modes, as expected. 

In a heterogeneous computing environment, code performance varies substantially with available resources, as seen by
contrasting corresponding results for (a,b,c) in Table \ref{tab:CostBreakdown}.      For
the slowest hardware,  the overall runtime will be only a factor few smaller than the corresponding CPU-only runtime.

\subsection{Binary neutron stars}
\begin{table*}
\begin{tabular}{lrr|ccccc|rr}
Version & srate & modes & $\tau_{start}$ & $\tau_{setup}$ & $\tau_{ad}$ & $\tau_{it,like}$ &$\tau_{it,rest}$ &
$\frac{T_{ILE}}{N_{eval}}$ & GPU \\  %
  &   Hz & m & sec & sec & & $\mu$sec & $\mu$sec  &sec  & use  \%\\ \hline 
CPU & 16384 & $\pm 2,\pm 1 $ & 35 & 26 & 0.14  & 680  & 25 & 590     \\ 
       & 4096 & $\pm 2, \pm 1$ & 35 & &  &   &&  25 &  \\ \hline
GPU (a) & 16384 & $\pm 2, \pm 1$   & 35 &  &&&  &    \\
       & 4096 & $ \pm 2, \pm 1 $        &  35  & &&&& 60--450  \\ \hline
GPU (b) & 16384 & $\pm 2, \pm 1$   & 35 & 26 & 0.07 &  2.4 & &35 & 16 \\
       & 4096 & $ \pm 2, \pm 1 $        &  35  & 11.5 & 0.1 &  2.4  & &24 &  \\ \hline
GPU (c) & 16384 & $\pm 2, \pm 1$   & 35 & 71 &&20 & 38 &  105   \\
       & 4096 & $ \pm 2, \pm 1 $        &  35  & 28 && 12 && 60  \\ \hline
\end{tabular}
\caption{\label{tab:CostBreakdown:BNS}\textbf{Profiling performance: Binary neutron stars}: Evaluation costs for the
  marginalized likelihood on default
  hardware,  analyzing $T=8\unit{s}$ of data with a binary neutron stars
  $m_1=1.4 M_\odot,M_2=1.35 M_\odot$, with convergence threshold $n_{\rm eff} > 50$.
}
\end{table*}

Table \ref{tab:CostBreakdown:BNS} shows the corresponding performance breakdown for our fiducial binary neutron star,
which has signal SNR=32.  In this analysis, we have adopted the TaylorT4 model with $\ell=2$ and all $m=\pm 2,\pm 1$ modes as our
fiducial analysis template.   This source has significantly longer duration and higher amplitude, both of which contribute to slower convergence and longer runtime. 

Low-mass compact binaries like binary neutron stars have much longer inspiral times from the same starting frequency of
$20\unit{Hz}$: roughly $160 \unit{s}$.   Nominally, the manipulation of a corresponding power-of-two duration data ($256\unit{s}$) involves
$32\times$ more costly Fourier transforms and time integrations than the BBH analysis described above.  
[An array of $256\unit{s}$ of data at $16\unit{kHz}$ corresponds to  32 Mb.]  Comparing $\tau_{setup}$ appearing in Tables
 \ref{tab:CostBreakdown} and  \ref{tab:CostBreakdown:BNS}, we indeed find our BNS analysis requires  a factor of order   $32\times$
 longer to set up all necessary inner products.  Unlike the BBH analysis, the inner product evaluation costs in
 $\tau_{setup}$ now dominates the
 overall evaluation time $T_{ILE}/N_{\rm eval}$.

Improved performance arises in part from re-using our adaptive integrator, which (after the first marginalized
likelihood evaluation) can exploit tight localization afforded by the many BNS cycles available in band and the high
amplitude of our fiducial BNS signal.   In some cases, the Monte Carlo integral converges to our target accuracy in of order $10\times $ fewer steps than for BBH.  
Additionally, particularly for first-stage RIFT grids, the improved performance arises from our choice of initial grid: many points fit poorly and identified as such well before the fiducial $2\times 10^6$ Monte Carlo iterations.  
Both sources of improved performance relative to the BBH analysis are independent of choice of hardware.

For the closed-form post-Newtonian approximation used in the study above, waveform costs are a modest contribution to
overall runtime, contributing only 4.1 sec to $\tau_{setup}$ at 16 kHz.
For other approximations not available in closed form, the waveform generation cost for binary neutron star inspirals
can be much more substantial, particularly as $f_{\rm min}$ decreases below the 20\unit{Hz} used in this study.

While GPU cards have finite memory, the modest size of our underlying data arrays and intemediate data products on-board ensures that we are unlikely to
saturate this bound, even for 16 kHz data, unless we investigate significantly lower starting frequencies or
employ vastly more angular modes $h_{lm}$.

\section{Performance and validation demonstrations with full pipeline}
\label{sec:end-to-end}

To better illustrate code performance in realistic settings, we also describe end-to-end analyses with
the original and modified code.     In a full analysis, many of the initially-proposed  evaluation points $\bm{\lambda}$ are
either not or marginally consistent with the data.  As a result, most marginalized likelihood evaluations in the first
iteration proceed extremely rapidly.  As the high-cost and low-cost evaluations are generally well-mixed between
different instances, the overall time to complete the full first iteration is typically much lower than subsequent
iterations.   With a well-positioned and sufficiently large initial grid,  few follow-on iterations are needed.  

\subsection{Binary black hole analysis}
\label{sec:sub:BBHFull}

\begin{figure*}
\includegraphics[width=0.45\textwidth]{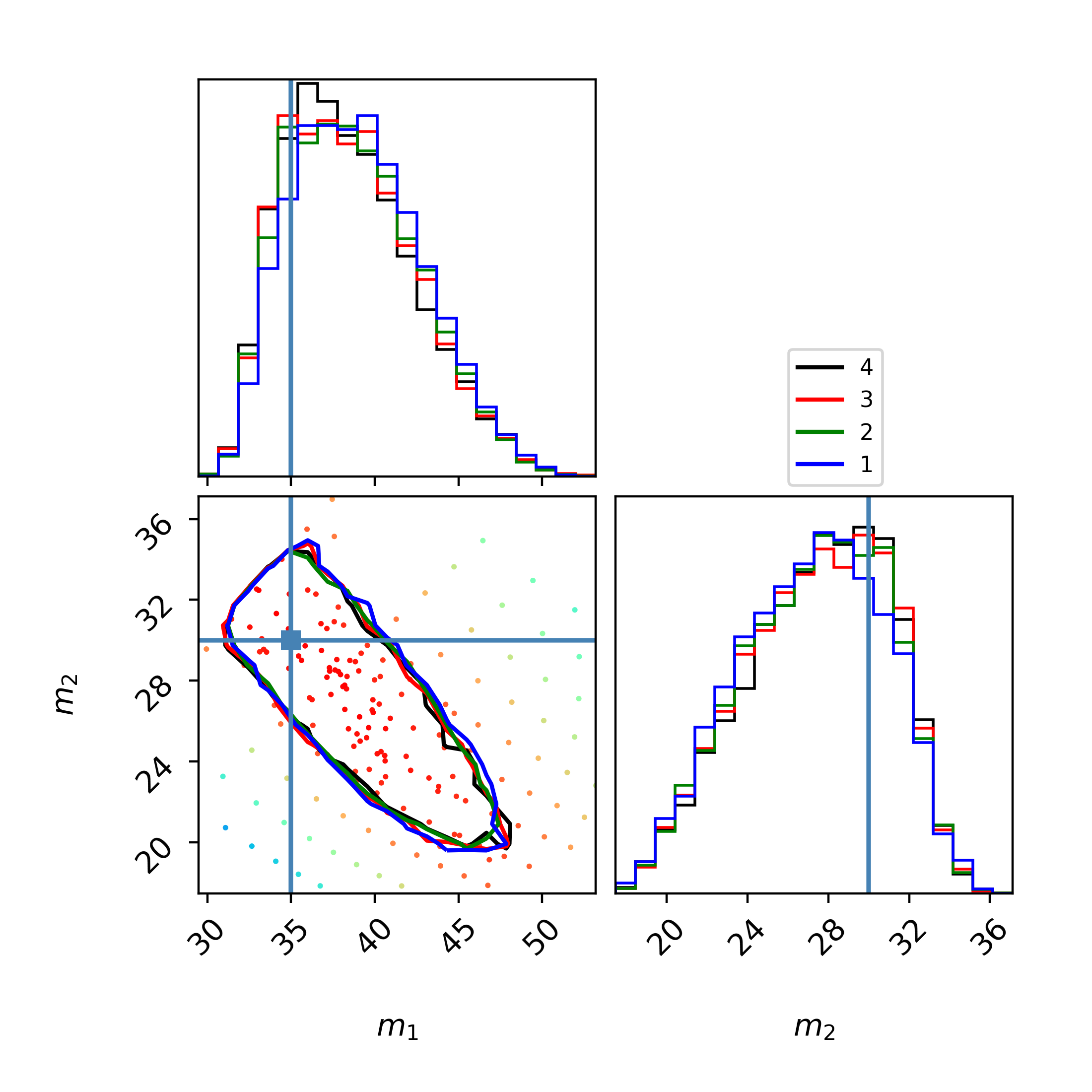}
\includegraphics[width=0.45\textwidth]{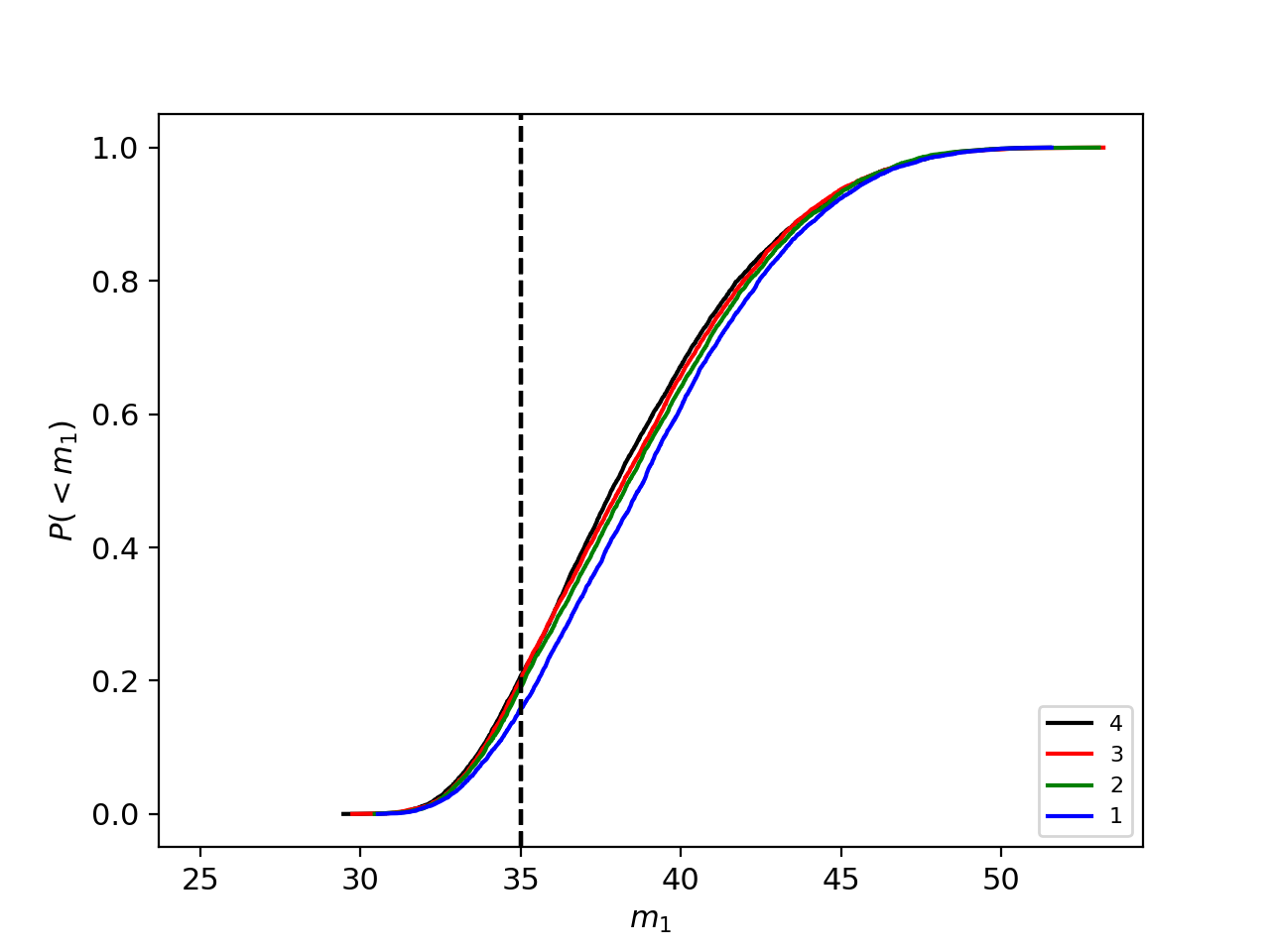}
\includegraphics[width=0.45\textwidth]{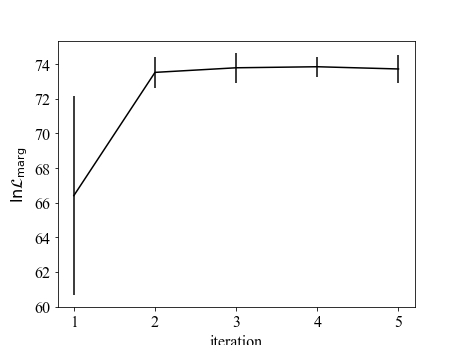}
\includegraphics[width=0.45\textwidth]{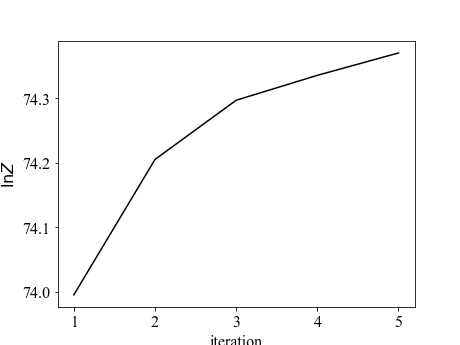}
\caption{\label{fig:BBH:MultiIterate}\textbf{Convergence of BBH analysis: Zero spin}: Results for marginal posterior distributions
  of our fiducial synthetic binary black hole.  Solid contours show credible intervals; solid one-dimensional distributions
  show marginal CDFs and PDFs for the corresponding variable; and colored points indicate the location $\bm{\lambda}$ and
  value of the underlying marginalized likelihood evaluations.   
\emph{Left panel } Posterior distribution
  over  $\mc$ and
  $\delta=(m_1-m_2)/M$.    \emph{Right panel}: Marginal 1d CDFs of $\mc$, showing convergence.
\emph{Bottom left}: Mean and variance of  \AddedResponse{the array $\ln{\cal L}_{\rm marg}(\bm{\lambda}_j)$  for
$j=1,2,\ldots N_{\rm eval}$ indexing all candidate sets of intrinsic parameters $\bm{\lambda}_j$ performed in that iteration},  showing that after the
first iteration the
candidate points are consistent with the posterior (i.e., no proposed point has very low $\ln {\cal L}_{\rm marg}$).
\emph{Bottom right panel}: The estimated evidence $Z = \int d\bm{\lambda} {\cal L}_{\rm marg}$ versus iteration number.  As systematic fitting error dominates our
error budget, Monte Carlo error is not shown.
}
\end{figure*}

Modest-amplitude short-duration binary black holes empirically constitute the most frequent detection candidates
for current ground-based GW observatories \cite{LIGO-O2-Catalog}.    Because of their brevity and hence broad posterior,
the RIFT code's interpolation-based method converges  rapidly.  Combined with the low cost of each iteration,
these sources require an exceptionally low committment of resources and can be performed in extremely low latency, as
desired.
To demonstrate this, we use the GPU-accelerated code with $N_{\rm eval}=20$, analyzing data with $f_{\rm
  sample}=4096\unit{Hz}$.    We use 100 points in the first iteration, in a very coarse mass  grid (i.e., spacing
comparable to typical astrophysical mass scales), followed by 20 points in each subsequent iteration.  Figure \ref{fig:BBH:MultiIterate} shows our results.   The
final posterior is already well-explored with the initial gridpoints, and converged by the second iteration.  
Conversely, on average all iterations required roughly 45 seconds per $\bm{\lambda}$ to evaluate their grid on GPU (a) hardware.  This
configuration therefore converged within the 30 minutes needed for the first two iterations.    We can achieve  smaller turnaround time for an otherwise identical analysis by adjusting $N_{\rm eval}$ appropriate to the available hardware.
While this low-dimensional problem does not capture all  fitting and automation challenges associated with high-dimensional
 fully precessing binaries, it does capture the low cost and rapid response possible with RIFT.

\subsection{Binary neutron star analysis : Assuming  zero spin}
\label{sec:sub:BNSFull}

\begin{figure*}
\includegraphics[width=0.45\textwidth]{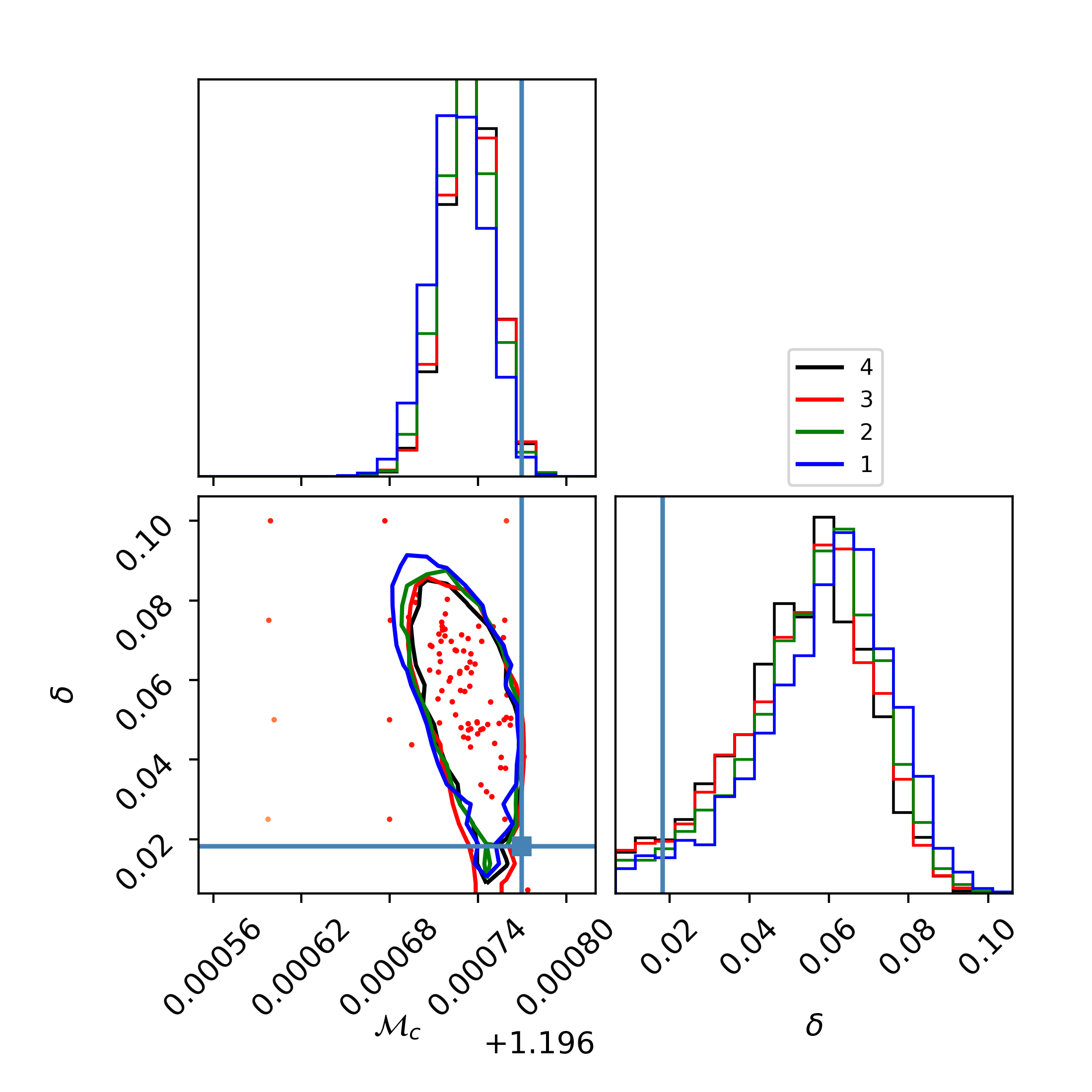}
\includegraphics[width=0.45\textwidth]{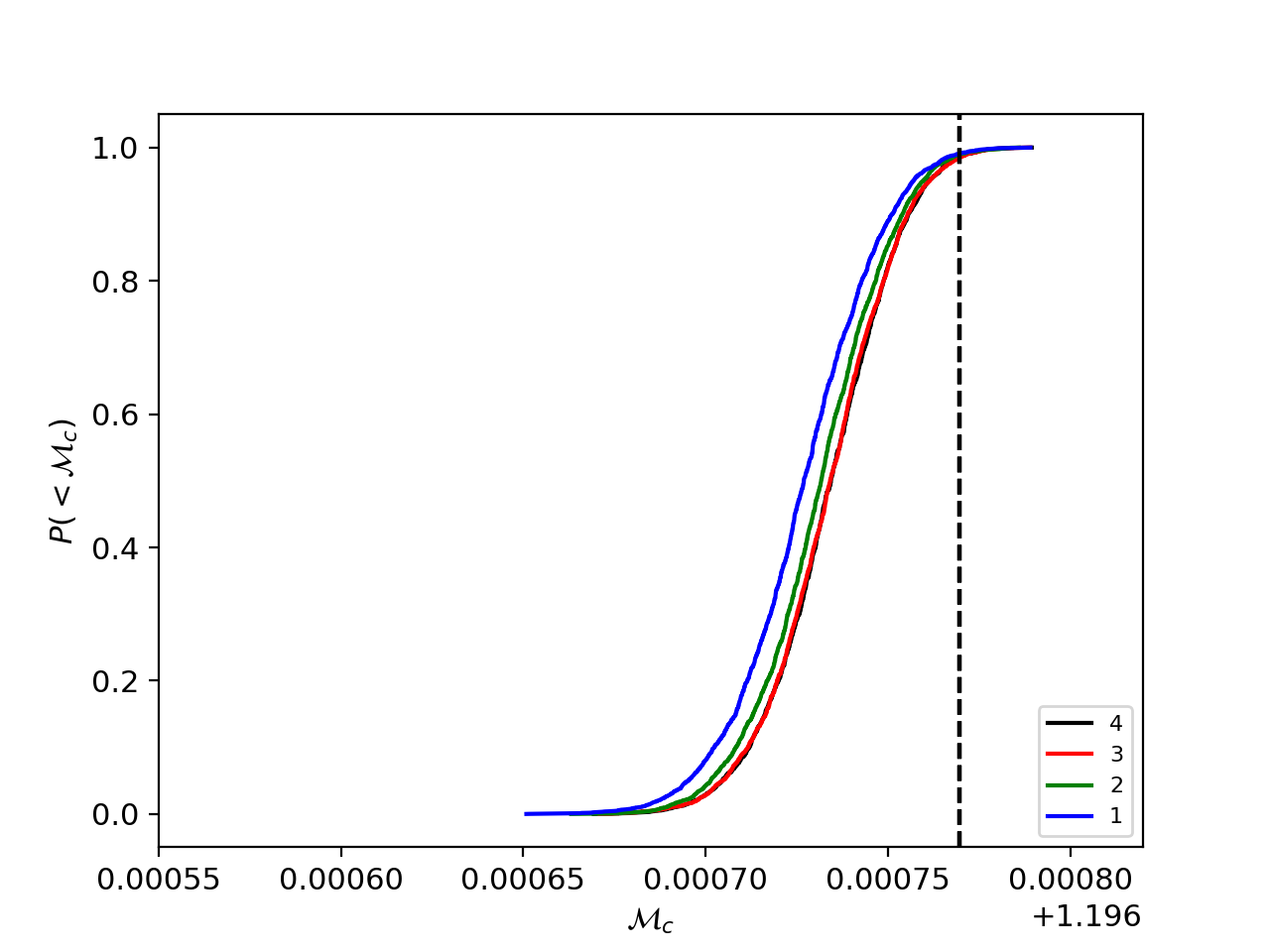}
\includegraphics[width=0.45\textwidth]{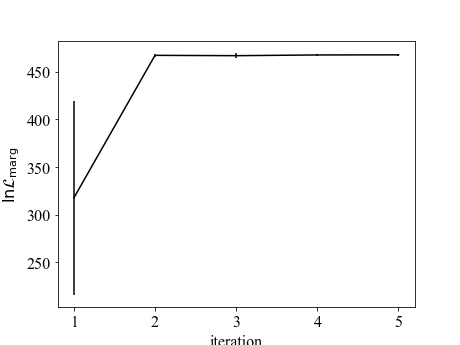}
\includegraphics[width=0.45\textwidth]{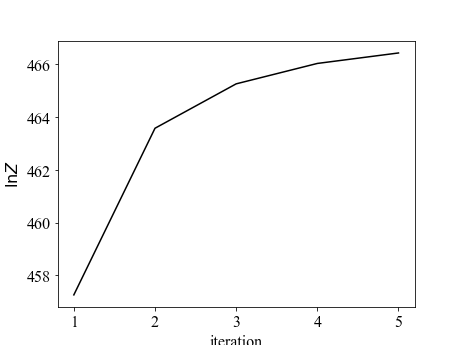}
\caption{\label{fig:BNS:MultiIterate}\textbf{Convergence of BNS analysis: Zero spin}: Results for marginal posterior distributions
  of our fiducial synthetic neutron star.  Solid contours show credible intervals; solid one-dimensional distributions
  show marginal CDFs and PDFs for the corresponding variable; and colored points indicate the location $\bm{\lambda}$ and
  value of the underlying marginalized likelihood evaluations.  
\emph{Left panel } Posterior distribution
  over  $\mc$ and
  $\delta=(m_1-m_2)/M$.    \emph{Right panel}: Marginal 1d CDFs of $\mc$, showing convergence.
\emph{Bottom left}: Mean and variance of  $\ln{\cal L}_{\rm marg}$ on the evaluation points,  showing that after the
first iteration the
candidate points are consistent with the posterior (i.e., no proposed point has very low $\ln {\cal L}_{\rm marg}$).
\emph{Bottom right panel}: $Z = \int d\bm{\lambda} {\cal L}_{\rm marg}$ versus iteration number.  As systematic fitting error dominates our
error budget early on, Monte Carlo error is not shown.
}
\end{figure*}

Significant-amplitude  binary neutron star mergers are the most important scenario for rapid parameter inference, as low
latency can enable multimessenger followup  \cite{LIGO-O2-Catalog}.    
In our second test we compare two workflows, one with the original embarrassingly-parallel RIFT code using $N_{\rm
  eval}=1$ and one with the
``batched'' GPU-accelerated code with $N_{\rm eval}=20$, analyzing data with $f_{\rm sample}=4096\unit{Hz}$.    
In this example, we use $100$ initial points spread over the two mass dimensions $\mc_z,\delta$, adding $20$ evaluations per
iteration.   We
choose this simple low-dimension, small-size, and slowly-converging configuration to facilitate visualization of the grid, posterior, and convergence.     The initial coarse grid covers a region $\mc_z\in[1.1962,1.1970] M_\odot$ and $\delta \in [0,0.25]$,  insuring the posterior was smaller than our initial
coarse grid spacing.     The top  panels of Figure  \ref{fig:BNS:MultiIterate} show posterior distributions derived
from the first several RIFT iterations, while the bottom panels show convergence diagnostics.
Because our analysis uses a post-Newtonian model to interpret a signal generated with SEOBNRv4, the peak posterior
density is slightly  offset
 from the synthetic signal's parameters

This specific workflow configuration  reduces overall core usage by maximizing GPU use per iteration: after the first
iteration, only one GPU is active.   
Specifically,  with the updated RIFT workflow used here, with one instance analyzing each 20 evaluations,  we use five
core+GPU pairs in the
first iteration, followed by one core+GPU for remaining iterations.
By contrast, with the original CPU-only RIFT code analyzing each $\bm{\lambda}$ in parallel, this process requires roughly 20 core-minutes
per $\bm{\lambda}$, using 100 cores in the first iteration and 20 cores in each
subsequent evaluation.   
Note that  because   $N_{\rm eval}=20$ for the GPU is larger than the (hardware-and $f_{\rm sample}$ dependent)  speedup factor between the CPU and
GPU implementation, the overall wallclock time needed for a end-to-end
analysis analysis is larger for the GPU workflow.
We can achieve comparable or smaller turnaround time for an otherwise identical analysis by adjusting $N_{\rm eval}$ appropriate to the available hardware.

\subsection{Binary neutron star analysis : Assuming  nonprecessing spin}
\label{sec:sub:BNS_spin}

Binary neutron star models with more parameters like spin and tides require correspondingly larger numbers of points in the
  initial grid and per iteration.  Fortunately, the number of observationally significant and accessible dimensions is often
  substantially less than the  prior  dimensionality.  In practice,we use roughly $20\times$ more points per
  iteration for precessing massive BBH systems ($d=8$) or for spinning binary neutron stars with tides ($d=6$).    As a
  result, we can still achieve relatively rapid turnaround on a high-dimensional binary neutron star analysis even in
  resource-constrained environments.

\begin{figure*}
\includegraphics[width=0.45\textwidth]{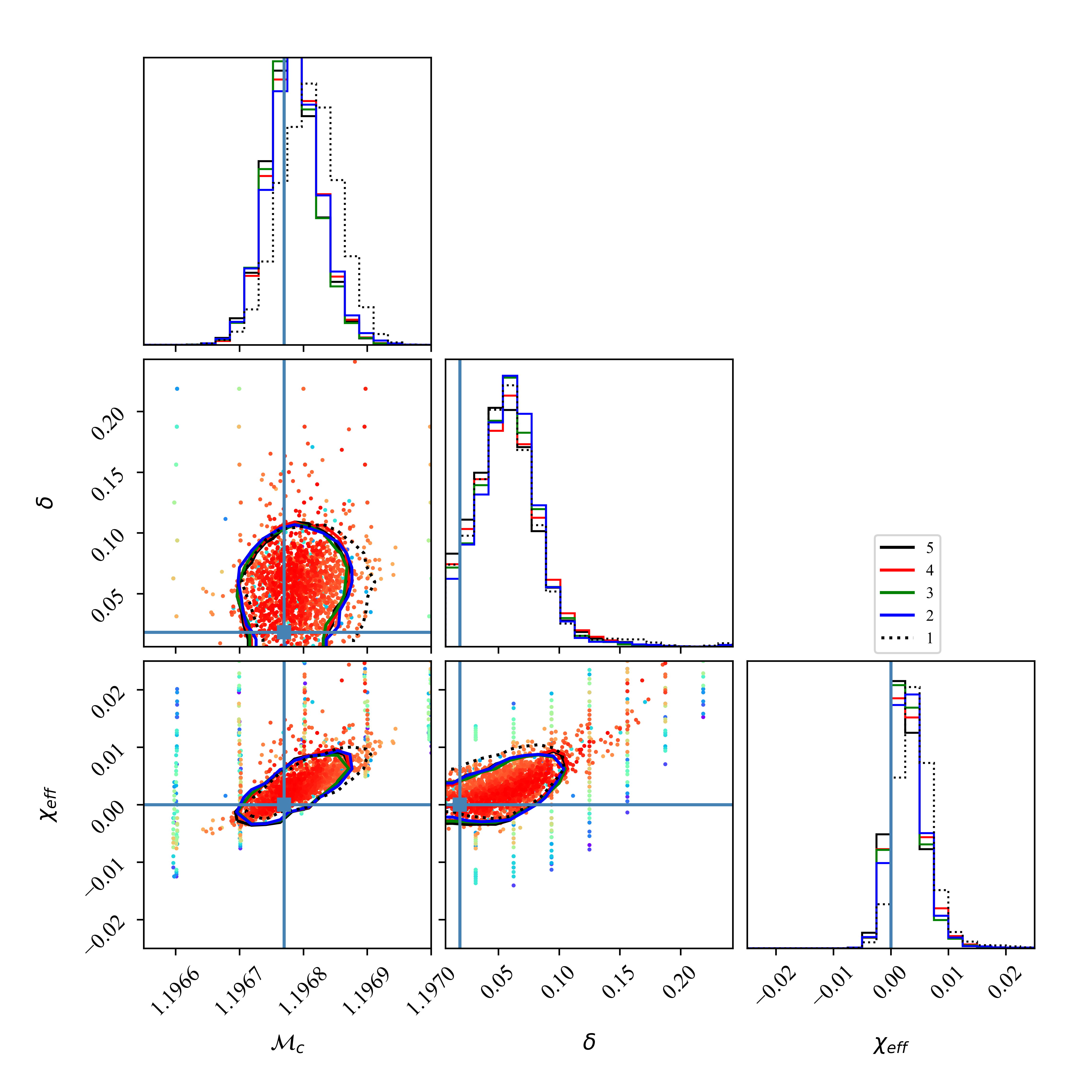}
\includegraphics[width=0.45\textwidth]{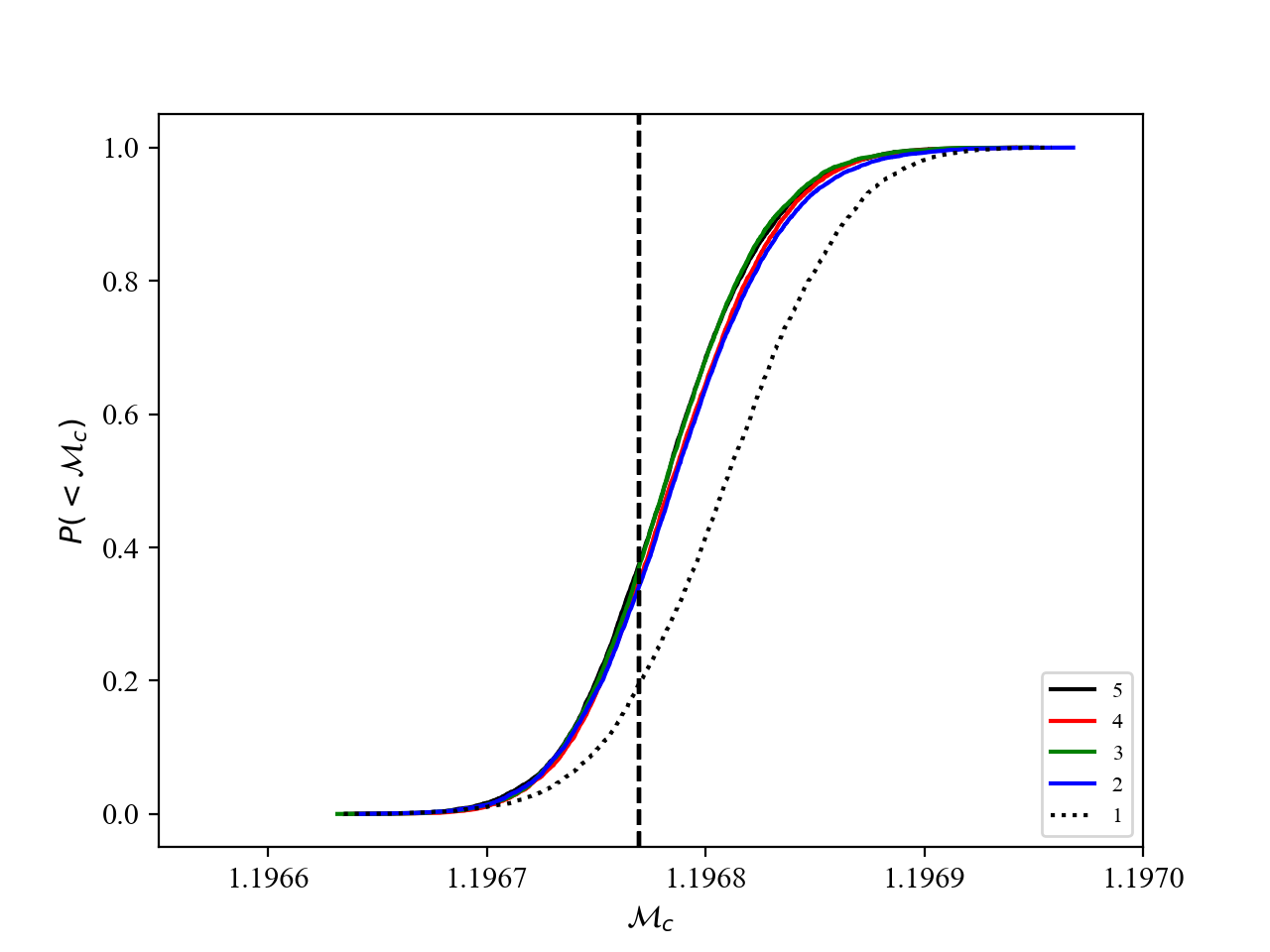}
\includegraphics[width=0.45\textwidth]{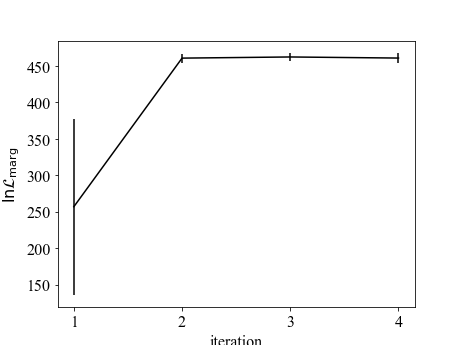}
\includegraphics[width=0.45\textwidth]{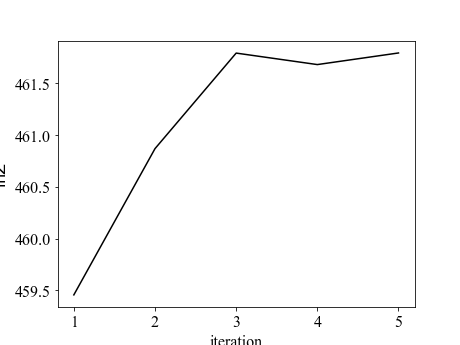}
\caption{\label{fig:BNS:Spin}\textbf{Demonstration of low-latency spinning analysis}:
Results for marginal posterior distributions
  of our fiducial synthetic neutron star, assuming the binary has nonprecessing spins.  Solid contours show credible intervals; solid one-dimensional distributions
  show marginal CDFs and PDFs for the corresponding variable; and colored points indicate the location $\bm{\lambda}$ and
  value of the underlying marginalized likelihood evaluations.  Only evaluations with marginalized log-likelihoods within $30$ of the maximum are shown, to increase contrast. The solid curves show an analysis with SEOBNRv4\_ROM.
 \emph{Left panel } Posterior distribution
  over  $\mc$,
  $\delta=(m_1-m_2)/M$, and $\chi_{\rm eff}$.    \emph{Right panel}: Marginal 1d CDFs of $\delta$, showing convergence.
\emph{Bottom left}: Mean and variance of  $\ln{\cal L}_{\rm marg}$ on the evaluation points,  showing that after the
first iteration the
candidate points are consistent with the posterior (i.e., no proposed point has very low $\ln {\cal L}_{\rm marg}$).
\emph{Bottom right panel}: $Z = \int d\bm{\lambda} {\cal L}_{\rm marg}$ versus iteration number.  As systematic fitting error dominates our
error budget early on, Monte Carlo error is not shown.
}
\end{figure*}

As a concrete demonstration of a realistic modest-latency analysis using the GPU-accelerated code, we reanalyze our fiducial binary
neutron star signal, accounting for the possibility of nonzero (aligned) neutron star spins.  Our initial grid consists
of 5000 points, spread \AddedResponse{approximately} uniformly across a 4-dimensional
hypercube $\mc_z\in[1.1962,1.1970] M_\odot$, $\delta \in[0,0.25]$ and $\chi_{i,z}\in[-0.05,0.05]$.  Subsequent
iterations use 500 random draws from the estimated posterior samples produced from the previous iteration.  
Because  our fiducial post-Newtonian model (TaylorT4) as implemented in \texttt{lalsuite}\cite{lalsuite} does not include spin, we
employ both the \texttt{SEOBNRv4\_ROM} and \texttt{SpinTaylorT4} waveform approximations for nonprecessing binaries,
omitting higher-order modes.   For SEOBNRv4\_ROM, we use 48 GPU (c) enabled nodes; for SpinTaylorT4, we use 100 GPU (a) nodes.
Figure \ref{fig:BNS:Spin} shows our results.

Due to our self-imposed resource constraints on $N_{\rm eval}=20$ and the number of GPUs, the first iteration is
resource-limited and requires
of order $(5000/20/100)\simeq
2.5$ to $(5000/20/48)\simeq
 5$ times as long as the first iteration of the zero-spin BNS analysis described above.  Each marginalized likelihood
 evaluation in the first iteration (as well as subsequent iterations) requires roughly 60 seconds for both SpinTaylorT4 and SEOBNRv4ROM, on average.   Subsequent
marginalized likelihood iterations are not resource-limited and complete in roughly 35 minutes, depending on hardware.
As previously, we can achieve comparable or smaller turnaround time for an otherwise identical analysis by adjusting $N_{\rm eval}$ appropriate to the available hardware.

\section{Conclusions}
\label{sec:conclude}
We have demonstrated that the marginalized likelihood $\ln {\cal L}_{\rm marg}(\bm{\lambda})$ appearing in the RIFT/rapidPE parameter
inference calculation can be evaluated at fixed $\bm{\lambda}$ in tens of seconds on average for
both binary black holes and binary neutron stars.  This performance improvement could enable very low latency source
parameter inference for compact binaries, which can be of use for targeting multimessenger followup observations via sky
localization and  precise source characterization.     This prospect is particularly interesting because RIFT can
often achieve this performance using computationally costly models for binary merger with rich physics like higher modes
or eccentricity, as the waveform generation cost does not usually limit code performance.   

\AddedResponse{In addition to producing results rapidly, RIFT results can be produced with a noticably smaller overall resource footprint than
  loosely similar LI analyses, even without tuning to optimize  RIFT pipeline settings.  As a concrete and non-optimized
  example,  for all five of the iterations of the spinning binary neutron star parameter inference with \texttt{SEOBNRv4\_ROM} from
  $20\unit{Hz}$ described in this work, our  RIFT analysis expended  roughly 14 core-days.  By contrast, a TaylorF2 analysis of GW170817
  with LI in MCMC mode starting from $23\unit{Hz}$ required 228 core-days.  %
  For precessing binary black holes using SEOBNRv3, the improvement is equally substantial: 10 core-days for a
  10-iteration investigation of a synthetic GW150914-like source with RIFT, versus 291 core-days for a LI analysis of
  GW170729.  
  We defer detailed relative benchmarking using comparably-converged parameter inference to future work.
}

The overall code performance and thus latency can be further decreased substantially, notably by converting the Monte
Carlo random number generation and inner products to GPU-based operations.  In such a configuration, the marginalized
likelihood code would perform almost all calculations (except waveform generation) on a GPU, with minimal communication off
the board.  This configuration should further reduce the average time needed to compute $\ln {\cal L}_{\rm
  marg}(\bm{\lambda})$ for both binary black holes (which are Monte Carlo limited) and binary neutron stars (which are
inner-product limited).   We anticipate a further factor of roughly 10 reduction in overall evaluation time can be
achieved soon.  At that level of performance, a single 8-GPU machine with contemporary hardware could perform parameter inference less than 10
minutes.    Since binary compact objects intersect our past light cone only once every roughly 15 minutes, accounting
for all past history, such a configuration would be able to address low-latency parameter inference for the duration of
2nd-generation ground-based observing.
Alternatively, if larger resource pools are available in low latency, both the original and now GPU-accelerated RIFT  can
perform extremely rapid parameter inference if large iterations are performed completely in parallel (i.e., $N_{\rm
  eval} \simeq 1$).

By allowing  models with higher-order modes to be used in low latency, our code can exploit the tighter constraints which higher modes can enable on  the properties of low-mass
binaries with suitable amplitudes and orientations.  These tighter constraints could
better  inform low-latency source classification and hence multimessenger followup observations of compact binary mergers.

Beyond low-latency multimessenger astronomy, rapid parameter inference enables new applications.  For example, every
parameter inference provide \emph{evidence} for a signal being present in the data; with rapid parameter inference, this
evidence could be used as (the last stage in a hierarchical pipeline for) a detection statistic
\cite{2018PhRvX...8b1019S}.    
This approach can identify individual events and
even a population.
Alternatively and in many ways equivalently, one can identify a population of GW sources without assuming any one is
real, by  applying parameter inference to more candidate events and self-consistently separating foreground and
background
\cite{2015PhRvD..91b3005F,2019MNRAS.tmp..230G}.

When suitable surrogate models are available, the overall code performance and thus latency could be yet again further
reduced by eliminating the iteration and fitting stages entirely, performing one Monte Carlo at once
\cite{gwastro-PE-AlternativeArchitecturesROM}.  This approach exploits a linear representation of  $h_{lm}(t)$  via
basis functions, to enable rapid likelihood evaluation as a function of both extrinsic parameters and  $\bm{\lambda}$.
Though not necessarily or compactly available for all surrogate models, particularly for the small basis sizes necessary
to fit onboard GPUs, this approach could enable exceedingly low latency at small computational cost.  This alternative
architecture would be exceptionally well-suited to the alternative applications of low-latency PE described above. 

The use of \texttt{cupy} enables our code to be highly portable across architectures and heterogeneous GPU environments,
while transitioning smoothly between GPU and CPU mode.     The techniques we used here will be transported to other
Bayesian inference modeling codes used to interpret GW observations \cite{gwastro-PopulationReconstruct-Parametric-Wysocki2018,gwastro-PopulationReconstruct-Code-PopModels}.

\AddedResponse{In this work, we have focused exclusively on profiling a simplified pipeline  to produce posterior distributions
  for detector-frame intrinsic parameters.   The  code  also produces
reliable extrinsic parameter distributions \cite{gwastro-PE-AlternativeArchitectures}.  With some postprocessing,
this pipeline  provides joint posterior distributions for all intrinsic and extrinsic parameter distributions together;
examples of these distributions have been published elsewhere \cite{LIGO-O2-Catalog}.  
Presently, rather than harvest extrinsic information from every iteration, we harvest joint intrinsic and
extrinsic information with a single final iteration, which we will implement shortly in our production pipeline.
}

\begin{acknowledgements}
We thank our anonymous referee for helpful feedback.
DW thanks the RIT COS and CCRG for support, and  NSF-1707965.
JL is supported by NSF PHY-1707965.  
RO'S is supported by NSF PHY-1707965 and PHY-1607520.
Y-LLF was supported in part by BNL LDRD projects No.\,17-029 and No.\,19-002, 
and New York State Urban Development Corporation, d/b/a Empire State Development, under contract No.\,AA289.
We thank the Computational Science Initiative at BNL for hosting GPU Hackathon 2018, during which a part of this work was conducted. We also thank Jeff Layton at NVIDIA for bringing CuPy to our attention.
\end{acknowledgements}

\appendix
\section{Properties of resources used}

\AddedResponse{
LIGO-CIT worker nodes, denoted by (a)  in the text, are principally S6 nodes with 2-CPU $\times 8$ core Opteron
2.3 GHz machines with 16 Gb of RAM, with GTX 1050 Ti cards with 4 Gb of RAM.  For LIGO-CIT, profiling reports reflect
performance averaged over the whole cluster and hence lacks the detailed reporting produced for the other configurations.
LIGO-LHO worker nodes with GPUs, denoted by (c) in the text, are a heterogeneous configuration mostly consisting of
(a).  Unlike profiling at LIGO-CIT, the profiling for LIGO-LHO reflects controlled tests on a single node.
The V100 machine (ldas-pcdev13, denoted by (b) in the text) is  a 24-core ES-2650 v4 machine with 4 GPUs, only one of
which is active in our tests: Tesla V100 with 16 Gb of RAM.
All non-GPU profiling was  also performed on this machine.  
}

\bibliography{paperexport,LIGO-publications}
\end{document}